\documentclass[%
reprint,
 amsmath,amssymb,
 aps,
prl,
floatfix,
]{revtex4-2}
\pdfoutput=1
\usepackage{graphicx}% Include figure files
\usepackage{dcolumn}% Align table columns on decimal point
\usepackage{bm}% bold math
\usepackage{siunitx}
\usepackage{xr}
\newcommand*\sref[1]{%
    \ref{S-#1}}

\begin{document}

\title{Molecular Simulations of Liquid Jet Explosions and Shock Waves Induced by X-Ray Free-Electron Lasers}% Force line breaks with \\

\author{Leonie Chatzimagas}
\author{Jochen S. Hub}%
 \email{jochen.hub@uni-saarland.de}
\affiliation{%
 Theoretical Physics and Center for Biophysics, Saarland University, Saarbrücken 66123, Germany
}%


\date{\today}% It is always \today, today,

\begin{abstract}
 X-ray free-electron lasers (XFELs) produce X-ray pulses with high brilliance and short pulse duration. These properties enable structural investigations of biomolecular nanocrystals, and they allow resolving the dynamics of biomolecules down to the femtosecond timescale. Liquid jets are widely used to deliver samples into the XFEL beam. The impact of the X-ray pulse leads to vaporization and explosion of the liquid jet, while the expanding gas triggers the formation of shock wave trains traveling along the jet, which may affect biomolecular samples before they have been probed. Here, we used molecular dynamics simulations to reveal the structural dynamics of shock waves after an X-ray impact. Analysis of the density in the jet revealed shock waves that form close to the explosion center, travel along the jet with supersonic velocities and decay exponentially with an attenuation length proportional to the jet diameter. A trailing shock wave formed after the first shock wave, similar to the shock wave trains in experiments. Although using purely classical models in the simulations, the resulting explosion geometry and shock wave dynamics closely resemble experimental findings, and they highlight the importance of atomistic details for modeling shock wave attenuation.
\end{abstract}

\maketitle

%\tableofcontents

\paragraph{Introduction.}

X-ray free-electron lasers (XFELs) are a source of X-ray radiation that enable novel experiments in the field of structural biology.
The high peak brilliance, short pulse duration and high repetition rates enable resolving the structure of biomolecular nanocrystals, also in a time-resolved manner \cite{chapman2019,pande,nogly,pandey,neutze,chapman2019, roedig, gisriel}.
%Due to the high peak brilliance of the X-ray pulses the biomolecular nanocrystals are destroyed after exposure. Thus, the samples are probed serially.
Liquid jets have been used to deliver the samples rapidly into the beam \cite{chapman2011,gruenbein,gruenbein2021}. 
Upon impact of the X-ray beam, not only the sample is destroyed, but also the segment of the liquid jet exposed to the X-ray pulse is vaporized. The expanding vapor drives the explosion of the liquid jet.

In recent years, several studies analyzed XFEL-induced explosions of liquid droplets or liquid jets \cite{stan,stan2016,blaj,ursescu,ganan, beyerlein} as well as the relevance of explosions for the design of crystallographic studies.
Stan \textit{et al.} used time-resolved imaging to study explosions in droplets and jets, revealing that the expanding vapor launches shock waves traveling across the drops or along the jet \cite{stan2016,stan}. In jets, the shock front may split, leading to a sequence of succeeding pressure and density oscillations. 
These findings have implications for the design of experiments performed at high X-ray repetition rate, for two reasons \cite{gruenbein, blaj}: 
First, the gap in the jet formed by the explosion should be replenished before the arrival of the next X-ray pulse, requiring either increased jet velocities or reduced X-ray pulse repetition rates.
Second, shock waves traveling backwards along the jet may pass the samples before they have been probed. The pressure and density oscillations may perturb the samples leading to lower crystallographic resolution.

The influence of shock waves on biomolecules has been studied both computationally and experimentally.
Molecular dynamics (MD) simulations modeling laser-induced shock waves revealed that hemoglobin is compressed during the passage of a shock wave, but the tetrameric structure remained intact \cite{wiederschein,voehringer}. Experimentally, comparing crystallographic data of hen egg-white lysozyme (HEWL) microcrystals collected from two succesive pulses with \SI{1.1}{\mega\hertz} repetition rate revealed no perturbation of the HEWL microcrystals \cite{gruenbein,wiedorn,yefanov}. However, since the jet diameter was smaller or approximately equal to the focal spot of the X-ray pulse, no shock waves might have been created according to descriptions by Blaj \textit{et al.} \cite{blaj}. In contrast, two recent studies found a degradation of diffraction data quality of HEWL and hemoglobin microcrystals as well as structural changes in the hemoglobin microcrystals when probing with an effective pulse repetition rate of \SI{4.5}{\mega\hertz} \cite{gruenbein2021,gruenbein2021haemoglobin}. Additional experimental and computational studies are required to clarify how the experimental design controls the formation and propagation of shock waves, and wether the shock waves affect the biological samples. Here, we used MD simulations to model jet explosions as well as the formation, dynamics, and attenuation of shock waves.

\paragraph{Methods.}
We developed a purely classical MD model of the liquid jet and its exposure to an XFEL pulse. We modeled the liquid jet as a water cylinder in vacuum using the all-atom SPC/E water model \cite{spce}, if not stated otherwise.
To test the effect of the water model and the importance of atomic details, additional simulations were carried out using the all-atom TIP4P/2005 water model \cite{abascal} or the coarse-grained MARTINI water model \cite{Marrink}.
We modeled the explosion of the liquid jet via instantaneous heating of the central jet segment with a Gaussian-shaped temperature distribution by assigning new velocities to the water molecules drawn from the Maxwell-Boltzmann distribution. 
Since we focussed on the shock wave dynamics, inside the liquid water, and not on the plasma dynamics, the photoelectric effect and Auger electrons were neglected.
We set up simulation systems with jet diameters between \SI{10}{nm} and \SI{80}{nm} hit by X-ray beams with intensity profiles with FWHM between \SI{1.5}{nm} and \SI{12}{nm} (Table \sref{tab:param}), where the jet diameter was always larger than the FWHM. For each system, we carried out 19 to 100 independent simulations in the microcanonical ensemble (Table \sref{tab:param}).
Details on the model and simulation parameters are presented in the Supplementary Material.

\paragraph{Liquid jet explosion and shock waves.}
\begin{figure}
\includegraphics{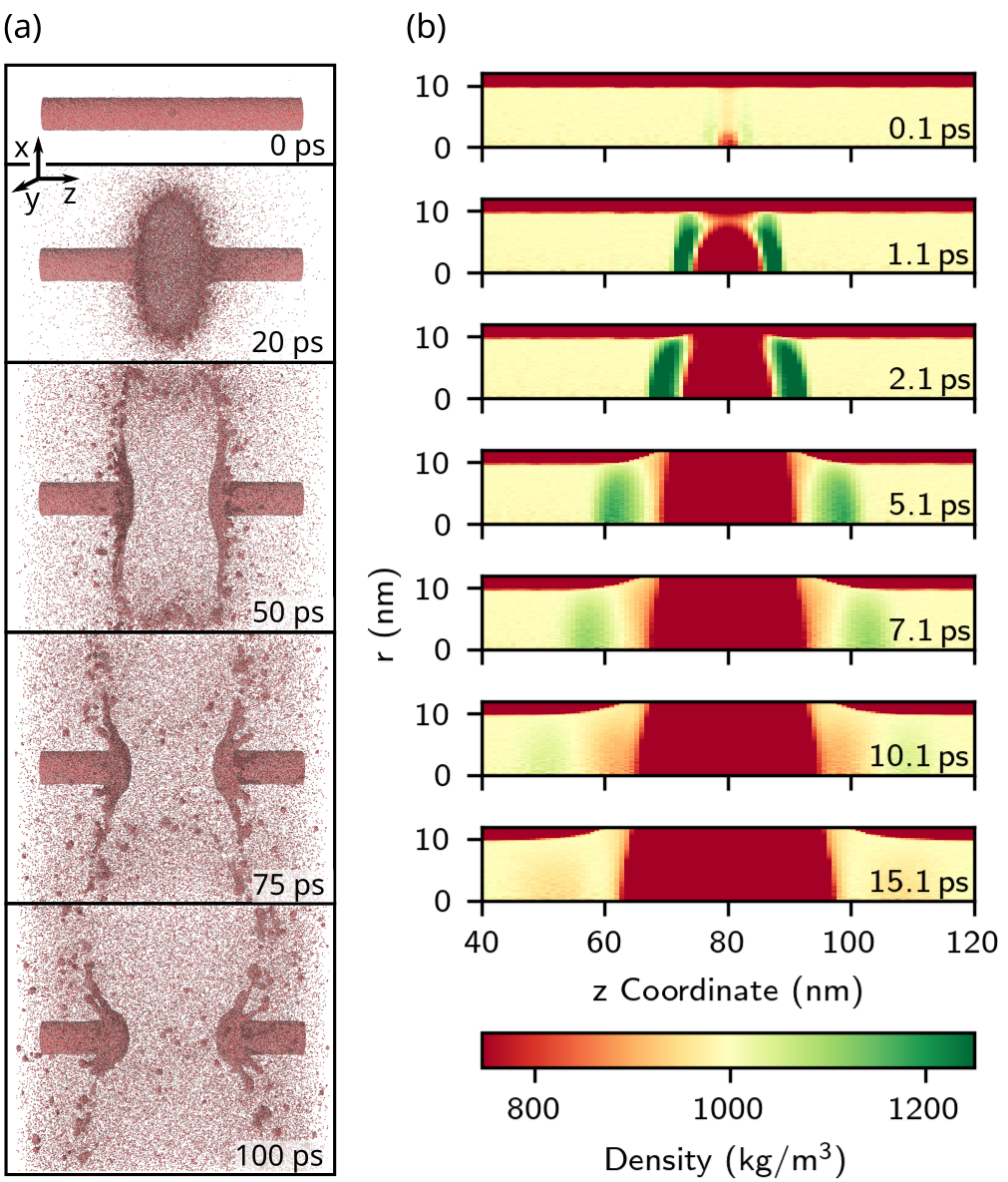}
\caption{Simulation of jet explosion with \SI{20}{\nano\meter} diameter induced by a modeled X-ray pulse with \SI{3}{\nano\meter} FWHM. (a) Snapshots of a single simulation. (b) Radial density at \SI{0.1}{\pico\second} to \SI{15.1}{\pico\second} after the X-ray impact averaged over 50 independent simulations. \label{jetcombi}}
\end{figure}

In the simulations, the impact of the X-ray pulse triggered an explosion of the central jet segment leading to the formation of a gap and, thereby, splitting the jet into two segments (Figs. \ref{jetcombi} (a) and \sref{jetcombiR5} (a)). As the gap growed over tens of picoseconds, thin water films formed at the ends of the segments, which later adopted a conical shape, and finally, after several tens of picoseconds, folded back to the jet. 
The simulated explosions developed on a by far smaller time scale compared to the experimental jet explosion observed by Stan \textit{et al.} \cite{stan} that developed in the microsecond range. 
The time scale difference is likely a result of the larger experimental jet diameter of \SI{20}{\micro\meter} diameter, which is approximately three orders of magnitude larger compared to our MD model. 
Despite these different length and time scales, the jet explosion dynamics in simulations resemble dynamics observed in the experiments.

To analyze shock waves traveling along the jet, we computed the mass density as a function of the axial and radial direction (Figs. \ref{jetcombi} (b) and \sref{jetcombiR5} (b), Supplementary Material Methods).
The data revealed the formation of two density peaks positioned symmetrically around the jet center and close to the explosion site. The peaks traveled along the jet and decayed to density values slightly below the density of an equilibrated water jet, revealing the rarefaction caused by the shock wave. Notably, the density peak was maximized near the jet axis (Fig. \ref{jetcombi} (b), $r=0$) but small or event absent at the jet surface ($r\approx 9$\,nm), suggesting that the shock wave energy is dissipated at the jet surface.

  \begin{figure}
  \includegraphics{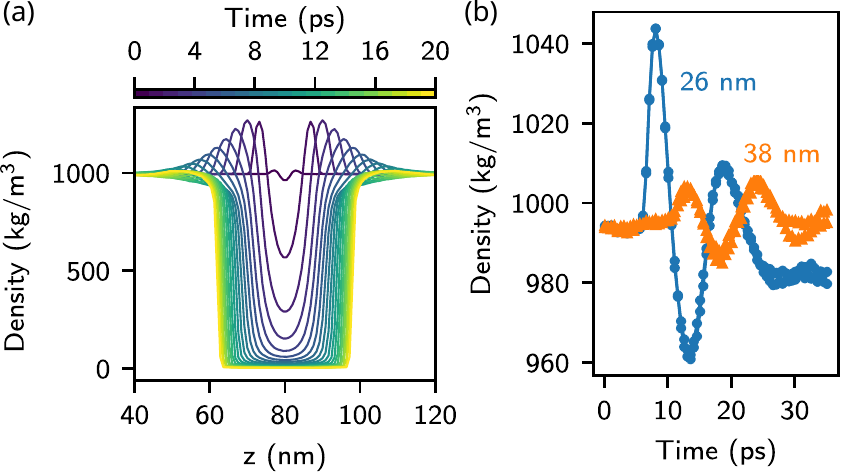}
  \caption{(a) Time evolution of the density in a water jet with \SI{20}{\nano\meter} diameter after explosion induced by a modeled X-ray pulse with \SI{3}{nm} FWHM, averaged over 50 simulations. Lines represent the density at time delays of 0.1 to \SI{20.1}{\pico\second} after X-ray impact (see color bar). (b) Density versus time at fixed distances of \SI{26}{\nano\meter} (blue) or \SI{36}{\nano\meter} (orange) from the jet center.   \label{densplot10}}
  \end{figure}

Averaging the two-dimensional densities along the radial direction yields the time-dependent density as function of the axial direction $z$. This analysis revealed a second density peak succeeding the first density peak (Fig. \ref{densplot10} (a)), as evident from the density evolution at fixed distances from the jet center (Fig. \ref{densplot10} (b)). 
All the simulations with different combinations of jet diameter and FWHM of the X-rays pulse showed an explosion of the jet and a similar evolution of the densities (Fig \sref{densall}).
Hence, despite (i) the approximations underlying the classical models and (ii) the smaller time and length scales, the MD simulations reproduce the sequential shock waves observed in previous experiments \cite{stan,blaj}.

\paragraph{Attenuation of the shock front.}
  \begin{figure*}
  \includegraphics{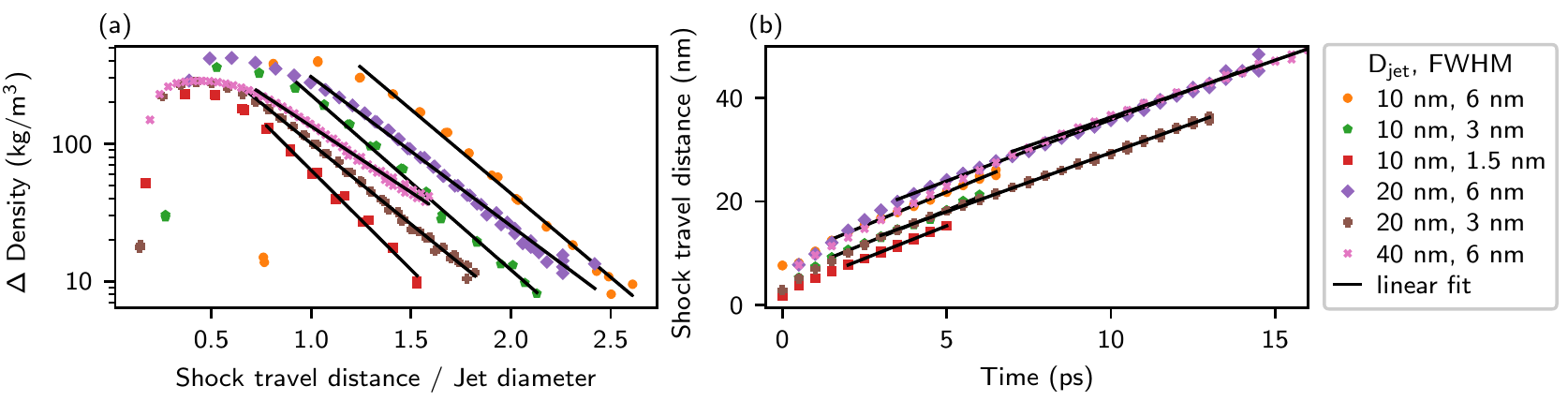}
  \caption{Shock wave attenuation and propagation. (a) Density peak attenuation of the shock front scaled by the water jet diameter. The relative height of the density peaks, $\Delta$ Density, attenuates with the propagation distance to the jet center. The linear attenuation on the semi-log plot indicates an exponential density decay. The attenuation approximately scales with the jet diameter. (b) Propagation of the shock front. The shock propagates with constant, supersonic velocity after few picosecond. This velocity is approximately the same for different simulation sets and in average $\sim \SI{2350}{\meter\per\second}$ (see Table \sref{tab:param}). \label{peakvelatten}}
  \end{figure*}
To analyze the attenuation of the shock front, we determined the height of the first density peak relative to the density of the equilibrated jet. 
Figure \ref{peakvelatten} (a) presents the height of the peaks for different jet diameters and X-ray pulse widths, plotted versus the propagation distance from the jet center. The attenuation of the density peaks followed an exponential decay, as evident from the linear decay on the semi-logarithmic scale. To test the influence of the jet diameter, we scaled the propagated distance of the density peaks by the jet diameter. A linear least-squares fit on the data revealed similar slopes for the peak decay in jets with different diameter, demonstrating that the decay length is proportional to the jet diameter or, equivalently, to the jet circumference (see Table \sref{tab:param}). These findings suggest that the shock wave energy is dissipated due to surface effects, compatible with the reduced shock wave densities at the surface (Figs. \ref{jetcombi} (b)).

These results are further compatible with experimental observations considering
(i) that, in experiments, the jets are three orders of magnitude larger and the shock front travels further into the jet compared to the simulations, reflecting that decay lengths increase with jet diameter; (ii) as pressure and density are positively correlated, the density evolution of the leading peak agrees with the pressure peak decay calculated from experiments by Blaj \textit{et al.} \cite{blaj}, who suggested that the pressure peaks decay exponentially and that the decay length is approximately proportional to the jet diameter. 

\paragraph{Shock wave velocity.}
Figure \ref{peakvelatten} (b) presents the peak propagation with time, demonstrating that, after a deceleration in the first \SI{1}{\pico\second} to \SI{4}{\pico\second}, the peaks propagate at a constant velocity after a few picoseconds. The shock wave velocities for all simulation systems is summarized in Table \sref{tab:param}, revealing velocities between \SI{2200}{\meter\per\second} and \SI{2600}{\meter\per\second}, where the velocities increase with smaller jets or with increased deposited X-ray energy. 
Hence, the density peaks travel with supersonic velocities, as expected for shock waves.

\paragraph{Evolution of the gap size.}
  \begin{figure}
  \includegraphics{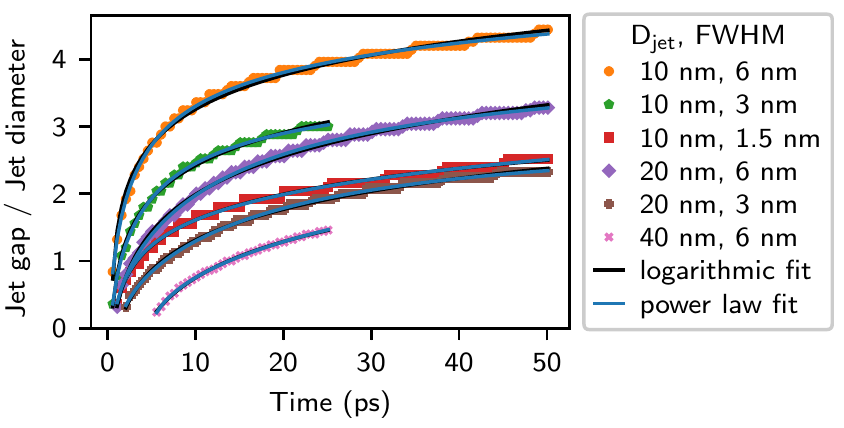}
  \caption{Evolution of the gap size. The gap was defined as the segment with a density below \SI{400}{\kilo\gram\per\cubic\meter}. The resulting gap size was scaled by the jet diameter and previously proposed models were fitted \cite{stan,blaj}. \label{gapgrowth}}
  \end{figure}

To analyze the gap growth induced by the jet explosion, we defined the gap as the axial segment with a density below \SI{400}{\kilo\gram\per\cubic\meter}, and we scaled the gap size by the jet diameter (Fig. \ref{gapgrowth}).
Notably, the gap size evolution in simulation resembled the evolution in experiments by Stan \textit{et al.} \cite{stan}.
However, in the simulations, the rise of the gap size was delayed by few picoseconds, as required for clearing the central segment after X-ray impact.
  
Two models have been put forward to describe the gap growth: a logarithmic growth used by Stan \textit{et al.} \cite{stan} and a power law proposed by Ga{\~n}{\'a}n-Calvo \cite{ganan}. 
To test wether these models are compatible with our simulations, we fitted the models to the simulated gap growth, augmented by a delay time before the initial rise of the gap (Supplementary Material Methods). 
As shown in Fig.\ \ref{gapgrowth}, both models are in excellent agreement with the data. However, since the simulation times-scales cover less then two orders of magnitude, the data is insuffcient to unambigiously determine all parameters in the power law model (see Table \sref{tab:paramganan}), or to decide whether one of the models is preferable for describing the gap growth. 
Taken together, the gap growth in simulations agrees with previous experiments and is compatible with the models by Stan \textit{et al.} \cite{stan} and Ga{\~n}{\'a}n-Calvo \cite{ganan}.

\paragraph{Influence of  water model and atomic details.}
To test the influence of the selected water model, we carried out additional simulations using the four-site TIP4P/2005 model.
The results obtained using TIP4P/2005 are in good agreement with results described above obtained with the three-site SPC/E water model (Fig. \sref{peakveldiff}, \sref{gapsizediff}), demonstrating that the choice of the atomistic water model is not critical for modeling shock wave propagation and gap growth.

\begin{figure}
  \includegraphics{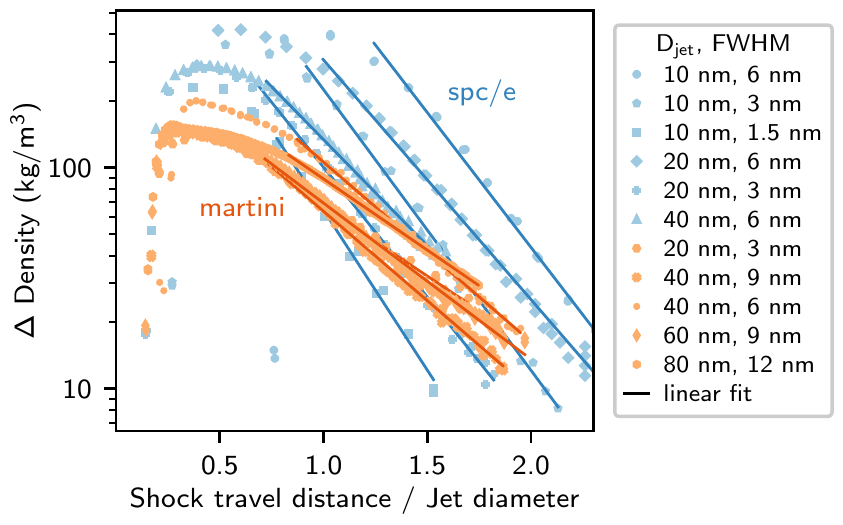}
  \caption{Influence of atomistic details on the density peak attenuation. The density decays over shorter distances when using the atomistic SPC/E model (blue) as compared to using the coarse grained MARTINI water model (orange).\label{martinicompare}}
  \end{figure}

To test the importance of atomic details for modeling jet explosion and shock wave propagation, we repeated the simulations with the coarse-grained MARTINI water model \cite{Marrink}. MARTINI describes liquid water as a Lennard-Jones fluid, modeling four molecules by one Lennard-Jones bead.
Hence, the MARTINI water model is greatly simplified as it lacks atomic details such as hydrogen bonds, leading to a smoothed potential energy landscape. Despite these simplifications, we observed qualitatively similar explosion and shock wave dynamics compared to simulations using the SPC/E water model (Fig. \ref{martinicompare}, Fig. \sref{peakvelmartini} and \sref{gapgrowthmartini}): the density peaks attenuate exponentially and propagate with a constant velocity after a few picoseconds; Further, the gap size evolution shows similar trends compared to experimental estimates \cite{stan}. 
However, the simulations reveal a quantitative difference of the shock wave attenuation dynamics compared to SPC/E simulations: While the exponential decay length is still approximately proportional to the jet diameter, the decay length scaled by the jet diameter is larger with MARTINI as compared to SPC/E simulations (see Fig \ref{martinicompare} and Table \sref{tab:param}).
We hypothesize that the slower decay is a consequence of the smoothed potential energy landscape of the MARTINI model, which may lead to reduced internal friction and, thereby, to slower dissipation of the shock wave energy. Hence, atomistic details are relevant for quantitatively describing the attenuation dynamics of the density peaks.

\paragraph{Conclusion.}
We carried out a large set of MD simulations of water jet explosions after impact of an XFEL laser pulse, modeled by a temperature jump of water molecules at the jet center. Despite the approximations underlying our classical simulations, we found good agreement with previous experiments in terms of explosion geometry, shock wave dynamics, and gap growth. These results suggests that photoelectric and Auger effects, which are certainly critical for modeling the explosion at the core of X-ray impact, are less important for shock wave formation and propagation or for the qualitative dynamics of the gap growth.
In the simulations, jet explosion triggered the formation of leading and trailing shock fronts that traveled with supersonic velocities along the jets. The shock waves attenuated exponentially with a decay length proportional to the jet diameter, suggesting a role of surface effects in dissipation of shock wave energy. Modeling the jet without atomic details led to slower decay of the shock wave, likely due to an overly smoothed potential energy landscape of water--water interactions. We expect these insights to be useful for designing experiments at XFELs with high repetition rate.

\begin{acknowledgments}
This project was supported by the Deutsche Forschungsgemeinschaft (HU 1971/3-1).
\end{acknowledgments}

%\nocite{*}
\bibliographystyle{apsrev4-2}
\bibliography{mybibabbr.bib}% Produces the bibliography via BibTeX.
\makeatletter\@input{supplementaux.tex}\makeatother
\end{document}

\title{Supplement: Molecular Simulations of Liquid Jet Explosions and Shock Waves Induced by X-Ray Free-Electron Lasers}% Force line breaks with \\

\author{Leonie Chatzimagas}
\author{Jochen S. Hub}%
 \email{jochen.hub@uni-saarland.de}
\affiliation{%
 Theoretical Physics and Center for Biophysics, Saarland University, Saarbrücken 66123, Germany
}%


\date{\today}% It is always \today, today,

\maketitle 
\section{ADDITIONAL FIGURES DISCUSSED IN THE MAIN TEXT}
  %
% Use \captionof and not a floating figure envirionment. This way, we avoid
% that the first page remains empty. Apparently, otherwise revtex hardcodes that 
% the first floating figure moves to page 2.
%
  \begin{figure}[h!]
\includegraphics{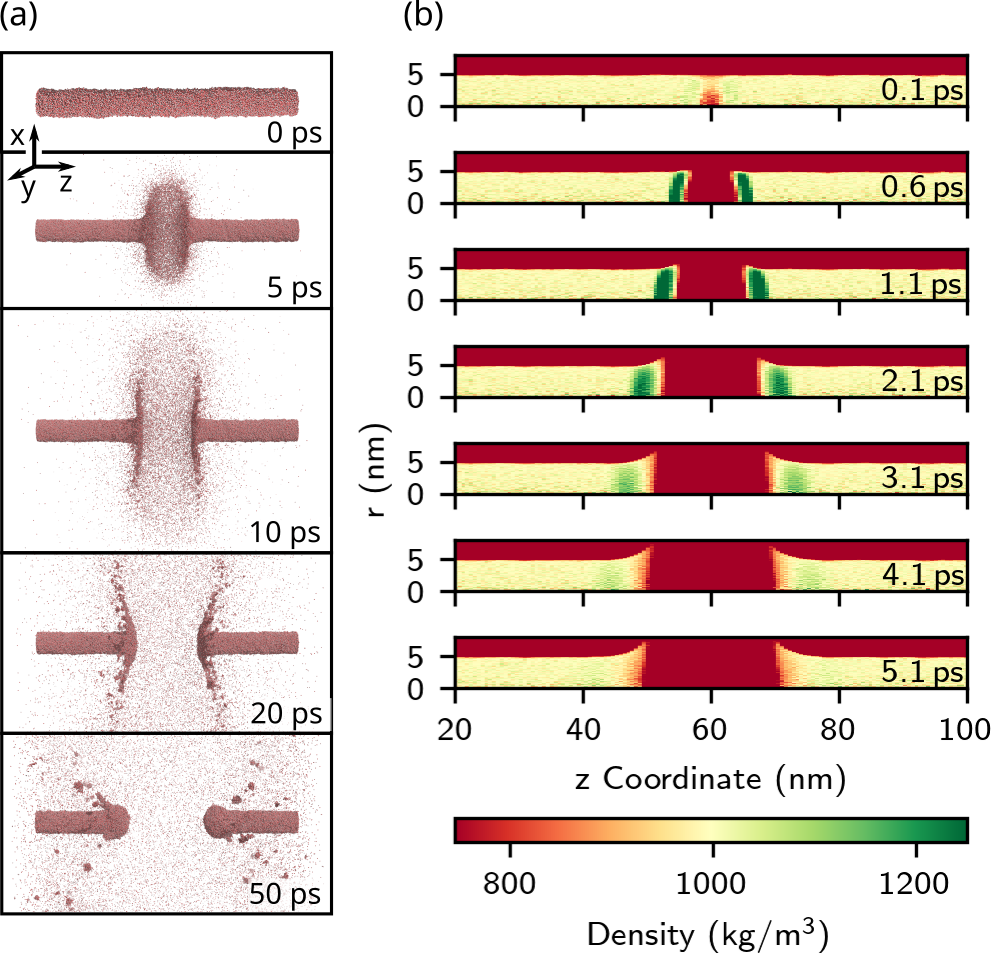}
\captionof{figure}{Simulation of the jet explosion in a water jet with \SI{5}{\nano\meter} diameter induced by a modeled X-ray pulse with \SI{3}{\nano\meter} FWHM. (a) Snapshots of a single simulation. (b) Radial density at \SI{0.1}{\pico\second} to \SI{5.1}{\pico\second} after the X-ray impact averaged over 100 independent simulations. \label{jetcombiR5}}
\end{figure}
  
  \begin{figure*}[h!]
  \includegraphics{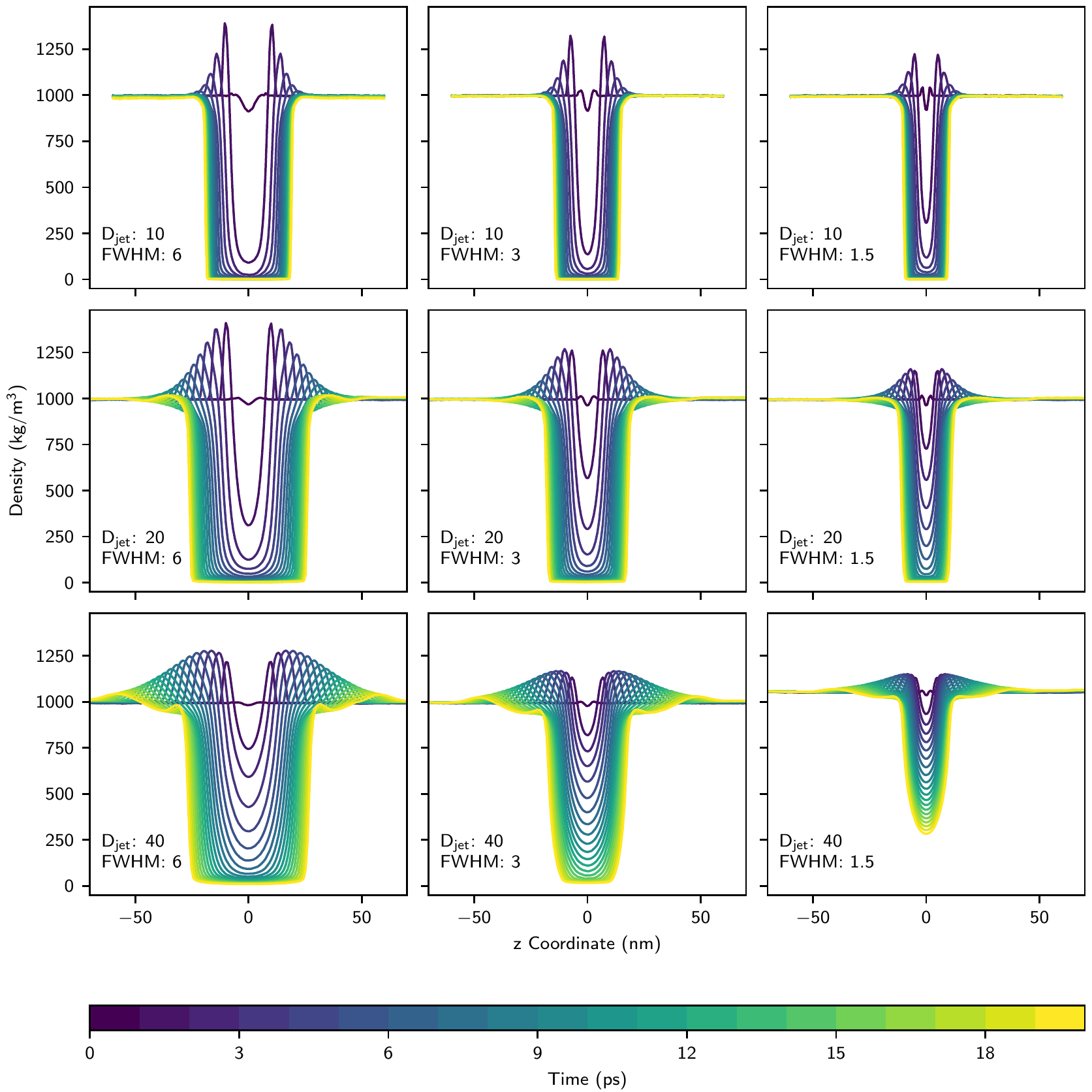}
  \captionof{figure}{Time evolution of the water density for simulations with different combinations of water jet diameter, D$_\mathrm{jet}$, and FWHM of the modeled X-rays pulse. D$_\mathrm{jet}$ and FWHM in nanometer are listed in panels. The densities reveal slower shock wave decay with larger D$_\mathrm{jet}$ and higher density peaks with increased FWHM, leading to increased deposited energy. For  D$_\mathrm{jet}=\SI{40}{\nano\meter}$ and $\mathrm{FWHM} = \SI{1.5}{\nano\meter}$, the deposited energy was insufficient for forming a gap (bottom right). \label{densall}}
  \end{figure*}

  \begin{figure*}[h!]
  \includegraphics{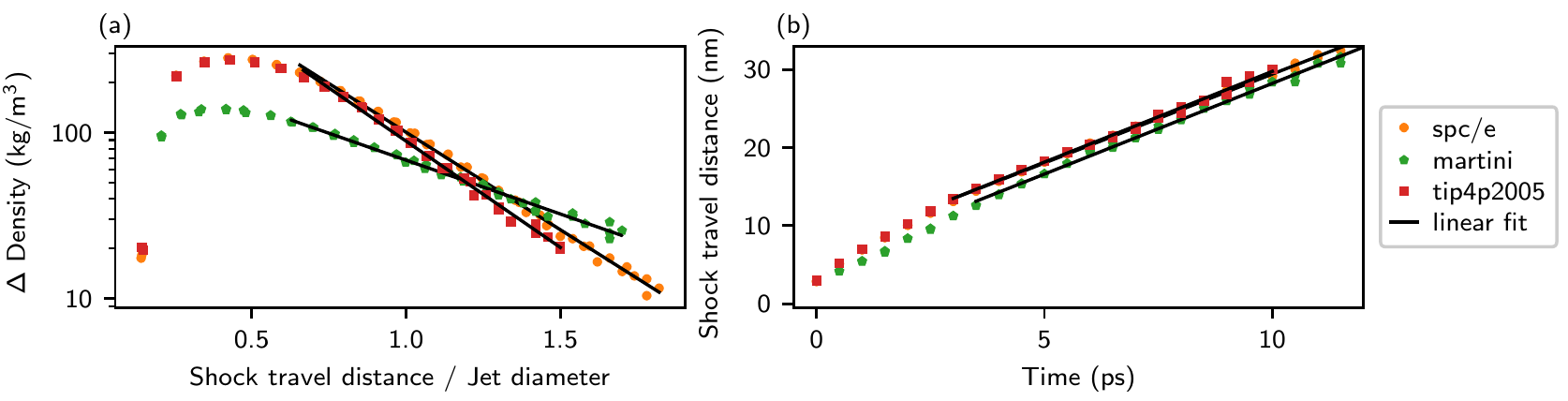}
  \captionof{figure}{Comparison of shock wave attenuation and propagation using the atomistic three-site SPC/E model (orange), the atomistic four-site TIP4P/2005 model (red), or the coarse-grained MARTINI water model (green), modeling the water as Lennard-Jones fluid. The densities were calculated from simulations of water jets with \SI{20}{\nano\meter} diameter after explosion induced by a modeled X-rays pulse with \SI{3}{\nano\meter} FWHM. Using different water models lead to qualitatively similar attenuation and propagation dynamics. (a) Density peak attenuation of the leading shock wave scaled by the water jet diameter. The relative height of the density peaks, $\Delta$ Density, attenuates with the propagation distance to the jet center. The linear attenuation on the semi-log plot indicates an exponential decay of the density. Results from TIP4P/2005 and SPC/E are nearly in quantitative agreement. In contrast, MARTINI leads to slower density decay. (b) Propagation of the first shock wave. The shock waves propagate with constant, supersonic velocity after few picoseconds regardless of the choice of the water model.\label{peakveldiff} }
  \end{figure*}

    \begin{figure}[h!]
  \includegraphics{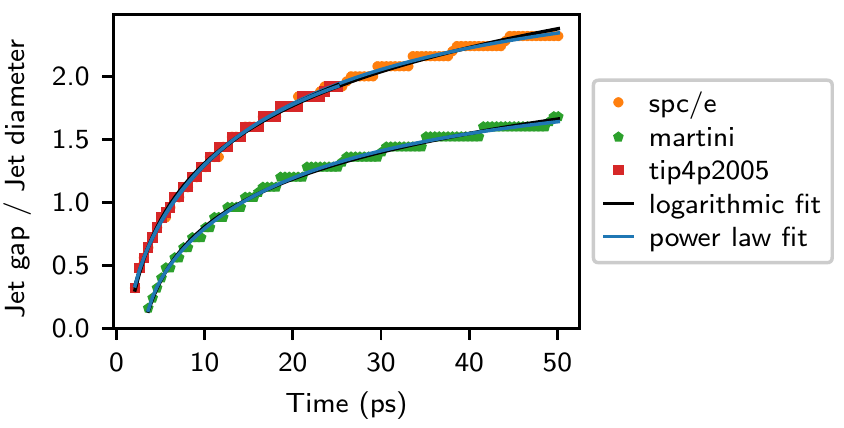}
  \captionof{figure}{Evolution of the gap sizes using the atomistic three-site SPC/E (orange) and four-site  TIP4P/2005 (red) water models as well as the coarse-grained MARTINI water model (green). The gap size was calculated from simulations of a water jet with 20 nm diameter after explosion induced by a modeled X-rays pulse with 3 nm FWHM. Different water models lead to qualitatively similar gap size evolution. However, while the two atomistic water models lead to nearly indistinguishable gap size evolutions, the MARTINI water model leads to a slower gap growth. Lines indicate fitted models of gap growth \cite{stan,blaj}.\label{gapsizediff} }
  \end{figure}

  \begin{figure*}[h!]
  \includegraphics{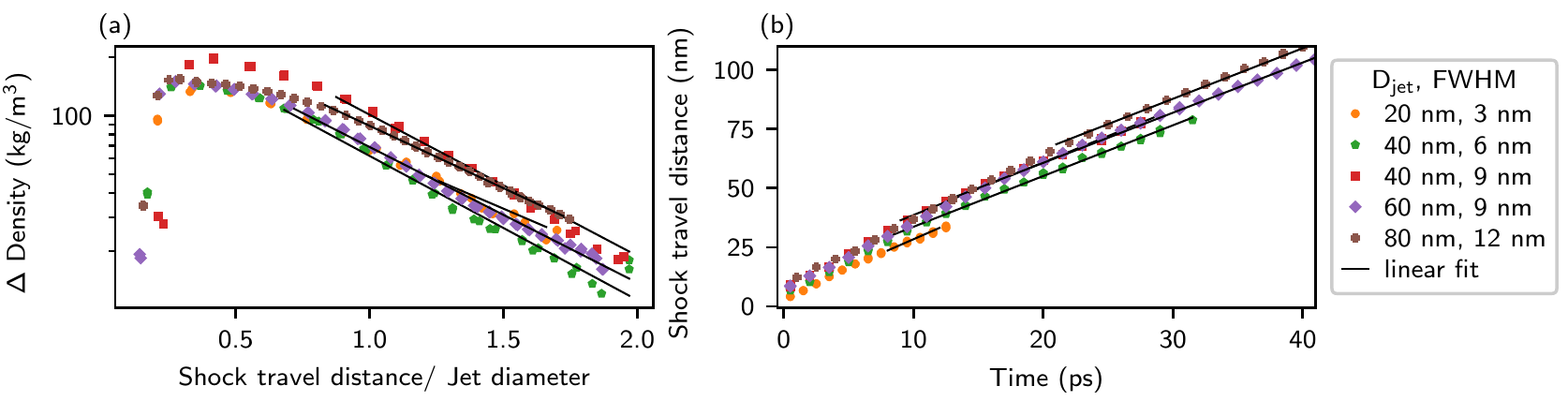}  \captionof{figure}{Shock wave attenuation and propagation using the MARTINI model. (a) Density peak attenuation of the leading shock wave scaled by the water jet diameter. The relative height of the density peaks, $\Delta$ Density, attenuates with the propagation distance to the jet center. The linear attenuation on the semi-log plot indicates an exponential density decay. The attenuation scales approximately with jet diameter. (b) Propagation of the first shock wave with constant, supersonic velocity of approximately $\SI{2210}{\meter\per\second}$ after few picoseconds (see also Table \sref{tab:param}).\label{peakvelmartini}}
  \end{figure*}
  
  \begin{figure}[h!]
  \includegraphics{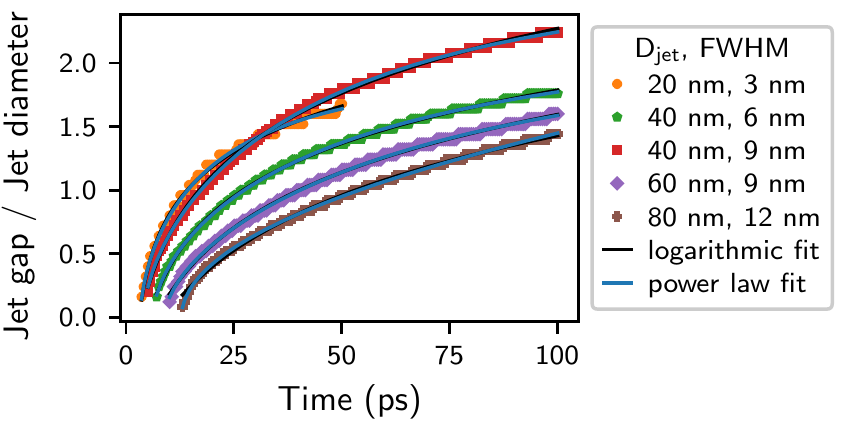}
  \captionof{figure}{Evolution of the gap size using the MARTINI model for systems with different jet diameters and X-ray pulse FWHM (colored symbols, see legend), qualitatively similar to results with SPC/E (cf.\ Fig. \mref{gapgrowth}). The resulting gap size was scaled by the jet diameter. Previously proposed models \cite{stan,blaj} were fitted to the data (lines).  \label{gapgrowthmartini}}
  \end{figure}

\clearpage
  
\section{SUPPLEMENTARY METHODS}
\subsection{Simulation system setup and heating of the jet}
The liquid jet was modeled as a cylinder of water molecules equilibrated at \SI{300}{\kelvin}. The jet was placed into a cubic box with periodic boundary conditions in a vacuum environment. Three water models were used to model the jet: the all-atom models SPC/E \cite{spce} and TIP4P/2005 \cite{abascal} and the coarse-grained (non-polarizable) MARTINI water model \cite{Marrink}. The geometries of atomistic water models were constrained with the SETTLE algorithm \cite{settle}.
The energy was minimized using a steepest descent algorithm, and the jet was equilibrated for \SI{25}{\pico\second} at a temperature of \SI{300}{\kelvin} using the velocity-rescaling thermostat \cite{bussi}.
The flow of the jet was not modeled whereas, in experiments, the jet travels with a velocity of typically \SIrange{10}{100}{\meter\per\second}. We neglected the jet flow in our simulations because the jet velocity is slow compared to the velocities of the shock waves.

We heated the central segment of the jet by assigning new velocities to the water atoms or beads at the jet center drawn from the Maxwell–Boltzmann distribution. Accordingly, new velocities were assigned to every atom $j$ within a cylinder that is orthogonally crossing the water jet. The cylinder radius was taken as $R = 3 \sigma$, where $\sigma$ is the standard deviation of a Gaussian-shaped intensity profile modeling the X-ray pulse.
The components $v_{i,j}$ of the new velocity of atom $j$ were drawn from the Maxwell–Boltzmann distribution,
 \begin{equation}\label{Boltzmann}
 P(v_{i,j}) = \sqrt{\frac{m_j}{2\pi k_bT(r)}} \exp\left( - \frac{m_j v_{i,j}^2}{2k_bT(r)} \right) \text{ ,}
\end{equation}
where $m_j$ is the mass of atom $j$ and $k_b$ the Boltzmann constant. The temperature $T(r)$ was taken as a Gaussian of distance $r$ of atom $j$ from the cylinder axis, orthogonal to the jet:
\begin{equation}
 T(r) = \left(T_{\mathrm{max}}-T_\mathrm{eq}\right)e^{- \frac{r^2}{2\sigma^2}} + T_\mathrm{eq}
 \label{gaussT}
\end{equation}
Hence, the temperature decayed from $T_\mathrm{max}= \SI{e5}{\kelvin}$ to the equilibrium temperature $T_\mathrm{eq}=\SI{300}{\kelvin}$. The maximum temperature $T_\mathrm{max}$ was chosen to match the plasma dynamics simulations by Beyerlein \textit{et al.} \cite{Beyerlein}.

\subsection{Simulation parameters for numerically stable NVE simulations}
Due to the high temperatures up to \SI{e5}{\kelvin} and, thus, large velocities, an integration timestep of \SI{0.02}{\femto\second} was required for obtaining numerically stable simulations with atomistic water models after heating the jet.
Because of the small integration timestep, double precision was required in GROMACS to prevent large energy drifts.
The decrease of the peak temperature during simulations enabled the use of larger integration time steps at later simulation times; namely, \SI{0.4}{\femto\second} for \SI{10}{nm} or \SI{20}{nm} diameter jets and \SI{0.2}{\femto\second} for \SI{40}{nm} diameter jets, respectively, after \SI{0.1}{\pico\second}.
In coarse-grained MARTINI water model simulations, the initial integration timestep was set to \SI{0.4}{\femto\second} and was increased to \SI{2}{\femto\second} after \SI{0.1}{\pico\second}, providing numerically stable simulations.

The methods for computing non-bonded interactions were critical for avoiding large energy drifts in the NVE simulations.
The Lennard-Jones interactions were computed using a cut-off with potential shift. The coulomb interactions were computed using the reaction field method with an infinite dielectric constant \cite{vanderspoel}. For both methods, the cut-off distance was set to 1.2 nm and the pair-lists were determined with the verlet cut-off scheme. While the particle-mesh Ewald \cite{pme} method likewise prevented an energy drift, it is inefficient for computing interactions in our systems due to the presence of large vacuum volumes.

The geometry of atomistic water molecules were constrained with the SETTLE algorithm to prevent large distortions of water binds and angles, which would lead to crashes in the simulation.

Statistically independent simulations were carried out by performing a \SI{5}{\pico\second} NVT simulation at \SI{300}{\kelvin} with a random set of initial velocities, followed by an independent heating of the jet center before each simulation run.

\subsection{Analysis}
For certain simulation sets, the pressure acting by the vaporized jet center was insufficient for splitting the jet into two segments. Such simulation sets were not considered in the analysis.

\paragraph{Density analysis.}
The density was computed in 200 bins along the jet axis and in 50 bins in radial direction. The density in each bin was averaged over the independent simulations. The number of independent simulations for each system is listed in Tab.\ \ref{tab:param}.

\paragraph{Density peak attenuation.}
To investigate the propagation of the first shock wave, we determined the density peak height in dependence of the distance of the peak position from the jet center for each time frame and simulation set. The density peak height was obtained by interpolating the density along the jet with a cubic spline. Then, two options were used to determine the density peaks: (i) the scipy tool signal.find\_peaks was used to find the position and height of the density peaks or (ii) the peak was defined as the maximum density value of the non-interpolated density data. Option (i) was used if possible; however, for larger times, the peaks were wide and noisy leading to over-fitting of the spline interpolation. Thus, if option (i) did not detect a peak, option (ii) was used. Furthermore, the peaks were only accepted if the density was larger than the thresholds of \SI{1005}{\kilo\gram\per\cubic\meter} and \SI{1050}{\kilo\gram\per\cubic\meter} for atomistic and MARTINI water models, respectivly. The height of the peak, $\Delta$ Density, was defined as the difference of the density at the peak relative to the averaged density of the equilibrated jet. 
\begin{table*}
\begin{ruledtabular}
\begin{tabular}{lllllllllll} 
Jet (nm)& Beam (nm) & water model & $N$ & $t$ (ps)& $v_\mathrm{S}$ (m/s)  & $\lambda_\mathrm{dec}$ (nm)& $m_\mathrm{log}$  \\ \hline
10 & 1.5 & SPC/E & 100 & 50 & \num{2521(2)}  & 33.3 & -3.3 \\
10 & 3 & SPC/E & 100 & 25 & \num{2592(2)}  & 29.2 & -2.9 \\
10 & 6 & SPC/E & 87 & 50 & \num{2601(2)}  & 28.1 & -2.8 \\
20 & 3 & SPC/E & 50 & 50 & 2279  & 54 & -2.7 \\
20 & 6 & SPC/E & 42 & 50 & 2368  & 49.8 & -2.5 \\
40 & 6 & SPC/E & 19 & 25 & 2201  & 88.2 & -2.2 \\
20 & 3 & TIP4P/2005 & 23 & 50 & \num{2325(1)} & 58.9 & -2.9 \\
20 & 3 & MARTINI & 25 & 100 & 2395 & 27.2 & -1.4 \\
40 & 6 & MARTINI & 20 & 100 & \num{2146(1)} & 73.7 & -1.8 \\
40 & 9 & MARTINI & 20 & 100 & \num{2227(1)} & 75.2 & -1.9 \\
60 & 9 & MARTINI & 20 & 100 & 2120 & 96.6 & -1.6 \\
80 & 12 & MARTINI & 20 & 100 & 2137 & 118.2 & -1.5 \\
\end{tabular}
\caption{Summary of the simulation parameters and results of the shock wave propagation: jet diameter ``Jet'', FWHM of the modeled X-ray beam ``Beam'', number of simulations $N$, simulation time $t$, velocity of the first shock wave $v_\mathrm{S}$, decay constant $\lambda_\mathrm{dec}$ and the slope of the log-linear fit to the scaled density decay $m_\mathrm{log}$ (\mref{peakvelatten}(a), \sref{peakveldiff}(a), \sref{peakvelmartini}(a)). Errors of $m_\mathrm{log}$ were below 2.5\%. \label{tab:param}}
\end{ruledtabular}
\end{table*}

\begin{table*}
\begin{ruledtabular}
\begin{tabular}{ccccccccccc}
Jet (nm) & Beam (nm) & water model  & $t_0$ (ps) & $\tau$ (ps) & $C$   \\ \hline
10 & 1.5 & SPC/E & 0.6 & 0.5 & 5.5  \\
10 & 3 & SPC/E & 0.4 & 0.3 & 7 \\
10 & 6 & SPC/E & 0.4 & 0.1 & 7.2   \\
20 & 3 & SPC/E & 1.3 & 1.4 & 13.4   \\
20 & 6 & SPC/E & 0.6 & 1 & 16.8 \\
40 & 6 & SPC/E & 4.2 & 3 & 28.2  \\
20 & 3 & TIP4P/2005 & 1.1 & 1.7 & 14.3 \\
20 & 3 & MARTINI & 3.1 & 1.7 & 9.9   \\
40 & 6 & MARTINI & 5.2 & 5.1 & 24.1   \\
40 & 9 & MARTINI & 3.3 & 4.9 & 29.9  \\
60 & 9 & MARTINI & 7 & 9.9 & 40.9 \\
80 & 12 & MARTINI & 9 & 15.5 & 59.4   \\
\end{tabular}
\caption{Estimated parameters of the logarithmic gap growth model, $t_0$, $\tau$ and $C$ according to Eq. \ref{gapfitstan}. ``Jet'' denotes jet diameter, and ``Beam'' denotes the FWHM of the modeled X-ray beam. Errors of $t_0$, $\tau$ and $C$ were below 1\%. \label{tab:paramstan}}
\end{ruledtabular}
\end{table*}

\begin{table*}
\begin{ruledtabular}
\begin{tabular}{ccccccccccc}
Jet (nm) & Beam (nm) & water model  & $t_0$ (ps)  & $C$ & $\tau_s$ (ps) & $\delta$  \\ \hline
10 & 1.5 & SPC/E & 0.8 & 0.7 & \num{3.6(7)} & 0.95 \\
10 & 3 & SPC/E & 0.4 & 1 & \num{6.3(6)} & 1.09\\
10 & 6 & SPC/E &  0.3 & 1.6 & \num{3.79(7)} & 1.07  \\
20 & 3 & SPC/E & 1.4 & 0.4 & \num{29(7)} & 1.18  \\
20 & 6 & SPC/E &  0.7 & 0.6 & \num{20(2)} & 1.13 \\
40 & 6 & SPC/E & 5 & 0.3 & \num{30(160)}  & \num{0.91(3)} \\
10 & 3 & TIP4P/2005 & 1.4 & 0.4 & \num{30(30)} & \num{1.22(18)} \\
20 & 3 & MARTINI & 3.3 & 0.3 & \num{32(17)} & 1.18  \\
40 & 6 & MARTINI & 6.2 & 0.2 & \num{82(24)} & 1.17  \\
40 & 9 & MARTINI & 4.1 & 0.2 & \num{92(15)} & 1.25  \\
60 & 9 & MARTINI & 9 & 0.1 & \num{170(60)} & 1.39 \\
80 & 12 & MARTINI & 12.8 & 0.2 & \num{0.05(15)} & \num{0.45(1)}  \\
\end{tabular}
\caption{Estimated parameters of the power law gap growth model, $t_0$, $C$, $\tau_s$, $\delta$, according to Eq. \ref{gapfitganan}. ``Jet'' denotes jet diameter, and ``Beam'' denotes the FWHM of the modeled X-ray beam. Errors of $C$ and $\delta$ were below 1\% if not reported otherwise. \label{tab:paramganan}}
\end{ruledtabular}
\end{table*}

\subsection{Models of gap growth}
We modified previously proposed models of gap growth to account for the delay time $t_0$ required for clearing the central jet segment after X-ray impact to a density below \SI{400}{\kilo\gram\per\cubic\meter}. Accordingly, the logarithmic growth model by Stan \textit{et al.} \cite{stan} was taken as 
 \begin{equation}
 X_\mathrm{retraction}/R_\mathrm{j} = C\ln(1+ (t-t_0)/\tau) \textrm{,}
 \label{gapfitstan}
\end{equation}
where $t_0$, $C$ and $\tau$ are fitting parameters and $R_\mathrm{j}$ is the radius of the jet. The power law by Ga{\~n}{\'a}n-Calvo \cite{ganan} was taken as
 \begin{align}
 X_\mathrm{retraction}/R_\mathrm{j} &= C (t-t_0)^{\alpha_0}  \left(1 + \left(\frac{t-t_0}{\tau_s}\right)^{\delta}\right)^{\left(\alpha_1 - \alpha_0\right)/\delta} \\
 \alpha_0 &=  2/(2+\gamma) \\
 \alpha_1 &= 5/3 - \gamma \textrm{ ,}
 \label{gapfitganan}
\end{align}
where $t_0$, $C$ and $\tau_s$ are the fitting parameters of the model and $\gamma = 1.5$ was choosen following Ga{\~n}{\'a}n-Calvo \cite{ganan}.

\bibliographystyle{apsrev4-2}
\bibliography{mybibabbr.bib}% Produces the bibliography via BibTeX.
\makeatletter\@input{mainaux.tex}\makeatother